# Exchange Bias in $BiFe_{0.8}Mn_{0.2}O_3$ Nanoparticles with an Antiferromagnetic Core and a Diluted Antiferromagnetic Shell

P. K. Manna and S. M. Yusuf*

*Solid State Physics Division, Bhabha Atomic Research Centre, Mumbai 400085, India*

R. Shukla and A. K. Tyagi

*Chemistry Division, Bhabha Atomic Research Centre, Mumbai 400085, India*

We have observed conventional signature of exchange bias (EB), in form of shift in field-cooled (FC) hysteresis loop, and training effect, in $BiFe_{0.8}Mn_{0.2}O_3$ nanoparticles. From neutron diffraction, thermoremanent magnetization and isothermoremanent magnetization measurements, the nanoparticles are found to be core-shell in nature, consisting of an antiferromagnetic (AFM) core, and a 2-dimensional diluted AFM (DAFF) shell with a net magnetization under a field. The analysis of the training effect data using the Binek's model shows that the observed loop shift arises entirely due to an interface exchange coupling between core and shell, and the intrinsic contribution of the DAFF shell to the total loop shift is zero. A significantly high value of EB field has been observed at room temperature. The present study is useful to understand the origin of EB in other DAFF-based systems as well.

# I. INTRODUCTION

The phenomenon of exchange bias (EB) has attracted a lot of attention both from theoretical and technological points of view [1]. The main indications of the presence of EB are normally identified as, (i) shift of field-cooled (FC) hysteresis loop along the magnetic field ($\mu_0 H$) axis, (ii) enhancement of coercivity ($\mu_0 H_C$) compared to the zero field cooled (ZFC) case, and (iii) presence of training effect (TE) *i.e.* a gradual decrease in EB field ($\mu_0 H_{eb}$) with increasing number of loop cycles ($n$) at a particular temperature [2]. Historically, EB was first observed in a ferromagnetic (FM) Co core and antiferromagnetic (AFM) CoO shell system [3]. Extensive research showed that this phenomenon can also be found in other systems *e.g.* FM/spin-glass, ferrimagnet (FI)/AFM, FI/FM, FI/FI [2] as well as bulk materials of phase-separated manganite [4] and cobaltite [5]. An interesting addition to these classes of materials is a system involving a diluted AFM (DAFF). Several reports [6-8] suggest that dilution of the AFM part by non-magnetic substitution (*e.g.* $Co_{1-x}Mg_xO$) or defects (*e.g.* $Co_{1-y}O$) can strongly influence EB properties of a FM/AFM system. Study of EB in AFM/DAFF system is scarce in literature. Benitez *et al.* [9-10] have recently reported the observation of shifted FC-hysteresis loop in $Co_3O_4$, CoO and $Cr_2O_3$ nanostructures having an AFM core-DAFF shell configuration. However, they [9-10] argued that since pure DAFF compounds (*e.g.* $Fe_{1-x}Zn_xF_2$) show a shift in FC-hysteresis loop because of the nucleation of metastable domain structures [11], the loop shift present in AFM core/DAFF shell type $Co_3O_4$, CoO and $Cr_2O_3$ nanostructures should not be considered as a signature of EB. On the other hand, Shi *et al.* [7] have termed the FC-loop shift phenomenon involving the same DAFF compounds $Fe_{1-x}Zn_xF_2$ (Co/$Fe_{1-x}Zn_xF_2$ bilayers) as EB. Several other reports on heterostructures involving DAFF compounds *e.g.* Co /$Fe_xNi_{1-x}F_2$ (DAFF) [8], Co/$Co_{1-x}Mg_xO$ (DAFF) [6], Co/$Co_{1-y}O$ (DAFF) [6], *etc.* have also termed such observation as a signature of EB. However, the point related to the FC-loop shift due to the intrinsic nature of DAFF was not addressed in references 6-8. From this discussion, it is clear that study of EB in an AFM core-DAFF shell is extremely important from fundamental point of view.

In this paper, we have studied EB in multiferroic $BiFe_{0.8}Mn_{0.2}O_3$ (BFMO) nanoparticles with an AFM core and a DAFF shell. The BFMO nanoparticles showed not only a shift in FC-hysteresis loop along $\mu_0 H$ as well as magnetization axes, but also TE phenomenon. By analyzing the TE data using the Binek's model [12], used in conventional EB-systems, we have shown that the observed FC-loop shift arises entirely due to an interface exchange coupling between core and shell, and the intrinsic contribution of the DAFF shell to the total loop shift is zero.

Besides this, the present work has other importance as well. The presence of novel magnetoelectric coupling between electric and magnetic orderings in such multiferroic materials allows one to take advantage of both magnetoelectric coupling, and interface exchange coupling (leading to EB in FM/multiferroic systems), in reducing writing energy of storage layer for magnetic electric random access memory (MERAM) [13]. Mn-substitution in $BiFeO_3$ (BFO) reduces leakage current and enhances magnetoelectric coupling at room temperature (RT) [14]. Moreover, being lead-free, it can replace lead-based material *viz.* lead zirconium titanate, which is currently being used in ferroelectric random access memory technology [15]. EB phenomenon is in the backbone of designing these magnetic memory elements. Some efforts have been made to understand the mechanism of EB in heterostructures involving BFO [16]. However, detailed study of the possible presence of EB in BFO itself is lacking. Tian *et al.* [17] reported EB phenomenon in polycrystalline $Bi_{1/3}Sr_{2/3}FeO_3$ compound in bulk form. However, EB field ($\mu_0 H_{eb}$) vanished at ~ 160 K, thereby limiting its application at RT [17]. In this paper, we report a significant value of $\mu_0 H_{eb}$ in the present multiferroic BFMO nanoparticles at RT.

Another interesting aspect of the present work is the observation of EB phenomenon without the conventional magnetic field cooling process. Conventionally, EB phenomenon appears in a coupled FM/AFM system, when it is field-cooled through Néel temperature ($T_N$) of the AFM material. However, in BFMO nanoparticles, we have found the FC-hysteresis loops to shift after cooling the nanoparticles from 310K, which is well below the $T_N$ of $BiFe_{0.8}Mn_{0.2}O_3$ (~560 K for its bulk form) [18] nanoparticles. Moreover, TE is also observed under the same condition.

## II. EXPERIMENTAL

The polycrystalline BFMO nanoparticles have been synthesized by the gel combustion method [19]. X-ray diffraction measurement was carried out at room temperature with a Philips x-ray diffractometer (X'pert PRO) using the monochromatized Cu-K$\alpha$ radiation. Transmission electron microscopy (TEM) images were recorded using a Philips CM30/Super TWIN Electron Microscope. Neutron powder diffraction measurement was carried out at 300 K at Dhruva reactor, Trombay, Mumbai, India using a five linear position sensitive detector (PSD) based powder diffractometer ($\lambda$ = 1.249 Å). The dc-magnetization measurements were carried out using a commercial vibrating sample magnetometer (Oxford Instruments). In the zero field cooled (ZFC) magnetization measurements, the sample was first cooled from 310 to 5 K in the absence of magnetic field and the magnetization was measured in the warming cycle under 0.05 T magnetic field. In the corresponding field-cooled (FC) magnetization measurements, the sample was cooled from 310 to 5 K in presence of the same magnetic field (as applied in the ZFC measurements) and magnetization was measured in the warming cycle by keeping the field on. In case of FC hysteresis measurements, the sample was cooled from 310 K to the required temperature under a desired magnetic field and hysteresis curves were recorded thereafter under ±9 T magnetic field whereas, the ZFC hysteresis loops were recorded after cooling the sample in zero field. To study TE [1, 12], the BFMO nanoparticles were first cooled from 310 to 5 K under a cooling field ($\mu_0 H_{cool}$) of 1T, and six consecutive hysteresis loops were recorded at 5 K. In the thermoremanent magnetization (TRM) measurements, [9] the sample was first cooled down from 310 to 5 K under a magnetic field. As the temperature reached 5 K, the applied field was switched off and the magnetization of the sample was measured. For the isothermoremanent magnetization (IRM) measurements, [9] the sample was cooled down to 5 K in the absence of magnetic field. After achieving 5 K temperature, the magnetic field was applied momentarily, removed again, and the magnetization of the sample was measured thereafter.

## III. RESULTS AND DISCUSSIONS

Rietveld refinement (using FullProf program [20]) of the x-ray diffraction pattern [Fig. 1 (a)] confirms the single phase nature of these nanoparticles which crystallize in a rhombohedral perovskite structure (space group: *R*3c) and the lattice constants were refined to be $a$=$b$=5.563(2), $c$=13.714(6) Å. Transmission electron microscopy (TEM) image reveals mean particle diameter to be 10-15 nm [Fig. 2(a)]. High resolution TEM (HRTEM) image [Fig. 2(b)] shows the crystalline nature in the core part and the presence of roughness/defects in surface part of the nanoparticles. Figure 3 shows the temperature dependence of ZFC and FC magnetization under $\mu_0 H$ = 0.05 T in the temperature range of 5-310 K. The bifurcation between ZFC and FC branches is present even up to 310 K, which signifies that the magnetic ordering temperature of the nanoparticles is higher than 310 K [21]. To check the microscopic nature of magnetic ordering, we have performed neutron diffraction experiment at 300 K. The neutron diffraction pattern (Fig. 4) has been fitted well (Rietveld refinement using FullProf program) [20] by using a model of G-type collinear AFM structure with $Fe^{3+}$/$Mn^{3+}$ magnetic moments [2.88(5) $\mu_B$ per Fe/Mn site at 300 K] oriented along the crystallographic $c$-axis. The oxygen octahedra were found to be tilted cooperatively from the $c$-axis by an angle of ~ 11.2º. In bulk form, BFMO compound orders antiferromagnetically at $T_N$~ 560 K [18]. With decrease in particle size, a reduction in $T_N$ is expected. However, behavior of the ZFC and FC curves shows that for the present nanoparticle system, a particle diameter of 10-15 nm is not enough to reduce the $T_N$ below 310 K. The splitting of the ZFC and FC curves below $T_N$ was also observed in core-shell type "AFM" $Co_3O_4$ nanowires [9] and it was attributed to the irreversible magnetization contribution arising from shell of the nanowires, which behaves like a 2-dimensional (2-D) DAFF. To check the presence of such shell for the present system, we have performed field dependence of thermoremanent (TRM) and isothermoremanent (IRM) magnetization measurements following the procedure employed in literature [9-10, 22]. A monotonically increasing TRM and an almost negligible value of IRM throughout the whole range of magnetic field [Fig. 5] signify the presence of a surface shell with a DAFF behavior [9-10]. The field dependence of TRM data has been fitted by the power law: TRM $\infty$ $(\mu_0 H)^{\lambda}$ [Fig 5], predicted theoretically for a 3-D random field Ising model [9-10]. For a 3-D DAFF system [9-10], $\lambda$ value was found to be greater than 1. However, for the present BFMO nanoparticles, the best fitted value of $\lambda$ was found to be 0.54±0.04 (<1). Benitez *et al.* [9-10] suggested that a 2-D DAFF system is likely to have a $\lambda$ value less than 1. Therefore, by combining the results of TRM/IRM, TEM and neutron diffraction studies, the present nanoparticles can be considered as core-shell type consisting of an AFM core and a 2-D DAFF shell with a net magnetization under magnetic field.

Magnetic nanoparticles with such core-shell morphology are potential system to study EB phenomenon. For this purpose, we have performed ZFC and FC hysteresis measurements after cooling the nanoparticles from 310K, which is well below the $T_N$ of BFMO (~560 K for its bulk form) [18] nanoparticles. A typical horizontal as well as vertical shift of the FC hysteresis loops (not present in the ZFC curve) is observed [Figs. 6(a) and 6(b)] along the negative magnetic field axis and positive magnetization axis, respectively, which can be found in a conventional EB system. It can be noted here that the amount of horizontal shift in FC-hysteresis loop under $\mu_0 H_{cool}$ = 5 T [Fig. 6 (b)] is greater than that at $\mu_0 H_{cool}$ = 1 T [Fig. 6(a)]. The amount of horizontal shift of the centre of FC-hysteresis loop is the measure of $\mu_0 H_{eb}$. The value of $\mu_0 H_C$ has been determined from half of the loop width. The cooling field ($\mu_0 H_{cool}$) dependence of $\mu_0 H_{eb}$ and $\mu_0 H_C$ at 5 K is plotted in figure 6(c). The magnitude of $\mu_0 H_{eb}$ increases with increasing $\mu_0 H_{cool}$ and showed a tendency of saturation at $\mu_0 H_{cool}$ = 7 T, whereas a monotonically increasing behavior of $\mu_0 H_C$ has been observed up to 7 T [Fig. 6(c)]. Similar $\mu_0 H_{cool}$ dependence of $\mu_0 H_{eb}$ was observed in core-shell type $Co_3O_4$ nanowires and it was attributed to an increase in frozen-in spins with increasing $\mu_0 H_{cool}$ [23]. The shift of the FC-hysteresis loop was evidenced at 300 K as well [inset of Fig. 6(a)], signifying the presence of exchange bias in the present BFMO nanoparticles even at room temperature. Temperature dependence of $\mu_0 H_{eb}$ and $\mu_0 H_C$ under $\mu_0 H_{cool}$ = 1 T is depicted in figure 7(a). A decrease in the magnitude of $\mu_0 H_{eb}$ was observed with increasing temperature, whereas $\mu_0 H_C$ showed an increasing tendency after an initial dip at ~50 K. The observed temperature dependence of $\mu_0 H_{eb}$ and $\mu_0 H_C$ is similar to that observed for $LaFeO_3$ (AFM) nanoparticles, where EB was observed after field cooling the sample from a temperature below its $T_N$ (similar to the present study) [21]. Ahmadvand *et al.* [21] explained it using the spontaneous EB mechanism [24]. The low temperature increase in $\mu_0 H_C$ seems to be correlated with the enhancement of FC-magnetization at low temperature [Fig. 3]. Spontaneous EB phenomenon has been discussed in literature both theoretically [24] and experimentally [21] and the reports suggest that it is possible (not an artifact of the experiment) [24] to induce EB in a FM/AFM system even when the AFM is cooled from a temperature less than its $T_N$. The observed vertical shift of the $M$ vs. $\mu_0 H$ curves along the positive magnetization axis is considered to be another important characteristic of an exchange coupled system [4, 21]. From the shift of the centre of $M$ vs. $\mu_0 H$ curves, we have obtained $M_{eb}$, which can be considered as vertical axis equivalent of $\mu_0 H_{eb}$ [4]. In fact, $\mu_0 H_{cool}$ and temperature dependence of $M_{eb}$ [Figs. 6(c) and 7(a), respectively] follow the same trend as that of $\mu_0 H_{eb}$. It is crucial to note that the amount of FC-loop shift, observed at RT for the present BFMO nanoparticles (~0.016 T under $\mu_0 H_{cool}$ =1T), is significantly higher than that reported in literature [= 0.00025 T for 14 nm (diameter) size $BiFeO_3$ nanoparticles] [25]. To explain EB phenomenon in FM/AFM heterostructures involving G-type AFM *e.g.* $BiFeO_3$, Dong *et al.* [26] proposed two mechanisms involving intrinsic Dzyaloshinskii-Moriya interaction and ferroelectric polarization. These two mechanisms are independent of the details of the FM spins and should be valid even in the presence of weak interface roughness. The only condition for the existence of these two mechanisms is the presence of oxygen octahedral tilting at the interface [26]. Recently, Borisevich *et al.* [27] have given a direct evidence for the presence of oxygen octahedral rotations across the interface of $BiFeO_3/La_{0.7}Sr_{0.3}MnO_3$ heterostructures using scanning transmission electron microscopy measurement. The present BFMO nanoparticles show a core-shell morphology because of the roughness/defects present at the surface, and a tilting of oxygen octahedra was also observed from the analysis of the neutron diffraction pattern (discussed earlier). Therefore, following the proposition of Dong *et al.* [26], an interface exchange coupling is expected between core and shell of the present BFMO nanoparticles, which can give rise to EB phenomenon.

To ascertain the presence of EB in the present nanoparticles, we have studied TE as well, which is considered to be an important characteristic of conventional EB-systems [1, 12]. The presence of TE in a FM/AFM system is a macroscopic fingerprint of deviation of AFM interface magnetization ($S_{AFM}$) away from its nonequilibrium configuration towards the equilibrium one during the field-cycling procedure [1, 12]. In short, TE originates due to training of $S_{AFM}$ and shift in FC-loop occurs because of an interface exchange coupling between $S_{AFM}$ and FM interface magnetization ($S_{FM}$) [12]. Based on this approach, Binek [1, 12] proposed a recursive formula for TE: $\mu_0 H_{eb}(n+1) - \mu_0 H_{eb}(n) = -\gamma[\mu_0 H_{eb}(n) - \mu_0 H_{eb}(\infty)]^3$, where $\mu_0 H_{eb}(n)$ and $\mu_0 H_{eb}(\infty)$ are magnitudes of EB field for *n*-th cycle and in the limit of infinite loops, respectively. $\gamma=1/(2\kappa^2)$, and $\kappa$ is a system dependent constant. Binek's formula has been applied successfully to fit the TE data of a wide variety of systems (*viz.* FM/AFM bilayers, FM hard/FM soft bilayers, FM nanodomains embedded in AFM matrix, spontaneously phase separated systems, double perovskite compound, core-shell nanoparticles, *etc*), where interface exchange coupling was the only origin of EB phenomenon. For the present BFMO nanoparticles, a monotonic decrease in $\mu_0 H_{eb}$ has been observed [Fig. 7(c)] with increasing *n*. Binek's model successfully fits [solid squares in Fig 7(c); the solid line is a guide to eye] the experimental data of BFMO nanoparticles signifying the validity of this model for an AFM core-DAFF shell system and the fitted parameters were found to be $\mu_0 H_{eb}(\infty)$=0.0877 T and $\gamma$= $1.0204\times10^4$ $T^{-2}$. Since, Binek's model is based on AFM and FM interface magnetization ($S_{AFM}$ and $S_{FM}$, respectively) [12], an excellent agreement of the experimentally observed TE data of the present BFMO nanoparticles with Binek's model clarifies that the observed shift in FC hysteresis loop occurs entirely due to an interface exchange coupling between core and shell, and the contribution of the DAFF shell alone to the total loop shift is zero. Analysis of the training effect data of BFMO nanoparticles thus, ensured that EB is indeed present in these nanoparticles, and the origin of EB phenomenon lies at the core-shell interface.

## IV. SUMMARY AND CONCLUSION

In summary, we have observed conventional signature of EB *viz.* shift in FC-hysteresis loop, and TE in multiferroic BFMO nanoparticles, even though the nanoparticles were field-cooled from a temperature lower than their $T_N$. The analysis of neutron diffraction, thermoremanent magnetization and isothermoremanent magnetization data shows that the nanoparticles consist of an AFM core, and a 2-dimensional DAFF shell having a net magnetization under field. Most importantly, by analyzing the TE data using the Binek's model, we have shown that the observed FC-loop shift arises entirely due to an interface exchange coupling between core and shell, and the contribution of the DAFF shell alone to the total loop shift is zero. A significantly high value of $\mu_0 H_{eb}$, observed at RT might have important implication in designing MERAM for its application at RT. The understanding gained in the present study would be of great help to throw light on the origin of EB in other EB-systems where DAFF forms one of their components.

* Corresponding author : smyusuf@barc.gov.in

**Figure captions**

FIG. 1. (color online) Observed (open circle) and Rietveld refined (solid line) x-ray diffraction pattern of $BiFe_{0.8}Mn_{0.2}O_3$ nanoparticles at room temperature. Solid lines at the bottom show the difference between observed and calculated patterns. Vertical lines show the positions of Bragg peaks. In the figure, x-axis has been plotted in terms of the magnitude of scattering vector Q [= $(4\pi/\lambda)\sin\theta$] where $\lambda$ is the wavelength of x-ray and $2\theta$ is the scattering angle. The most prominent peaks are indexed.

FIG. 2. (color online) (a) TEM image showing morphology of the $BiFe_{0.8}Mn_{0.2}O_3$ nanoparticles, (b) HRTEM image revealing surface defects/roughness.

FIG. 3. (color online) (a) Temperature dependence of ZFC and FC magnetizations under a magnetic field of 0.05 T.

FIG. 4. (color online) Rietveld refined neutron diffraction pattern of $BiFe_{0.8}Mn_{0.2}O_3$ nanoparticles at 300K, showing the presence of antiferromagnetic Bragg peak (indicated by arrow). Open circle and solid line indicate the observed and the calculated patterns, respectively. Solid lines at the bottom show the difference between observed and calculated patterns. Vertical lines show the positions of Bragg peaks. In the figure, x-axis has been plotted in terms of the magnitude of scattering vector Q [= $(4\pi/\lambda)\sin\theta$].

FIG. 5. (color online) Field dependence ($\mu_0 H$) of TRM and IRM to establish the core-shell nature of the nanoparticles. The solid line indicates the fitting of the TRM data by using the power law: TRM $\propto (\mu_0 H)^{\lambda}$, while the dotted line joining the IRM data is a guide to eye.

FIG. 6. (color online) (a) Zero field cooled (ZFC), and field-cooled (FC) $M$ vs. $\mu_0 H$ curves at 5 K. The FC curve was recorded under 1 T cooling field. Inset shows the shift of the FC-loop at 300 K under 1 T cooling field, (b) Zero field cooled (ZFC), and field-cooled (FC) $M$ vs. $\mu_0 H$ curve at 5

K, where the FC curve was recorded under 5 T cooling field, (c) Cooling field ($\mu_0 H_{cool}$) dependence of negative EB field ($-\mu_0 H_{eb}$), coercivity ($\mu_0 H_C$) and vertical loop shift ($M_{eb}$).

FIG. 7. (color online) (a) Temperature ($T$) dependence of $-\mu_0 H_{eb}$, $\mu_0 H_C$ and $M_{eb}$, (b) EB field ($-\mu_0 H_{eb}$: open circle) dependence on the number of field cycles ($n$). The solid squares represent the calculated data points using Binek's recursive formula, and the solid line is a guide to eye.

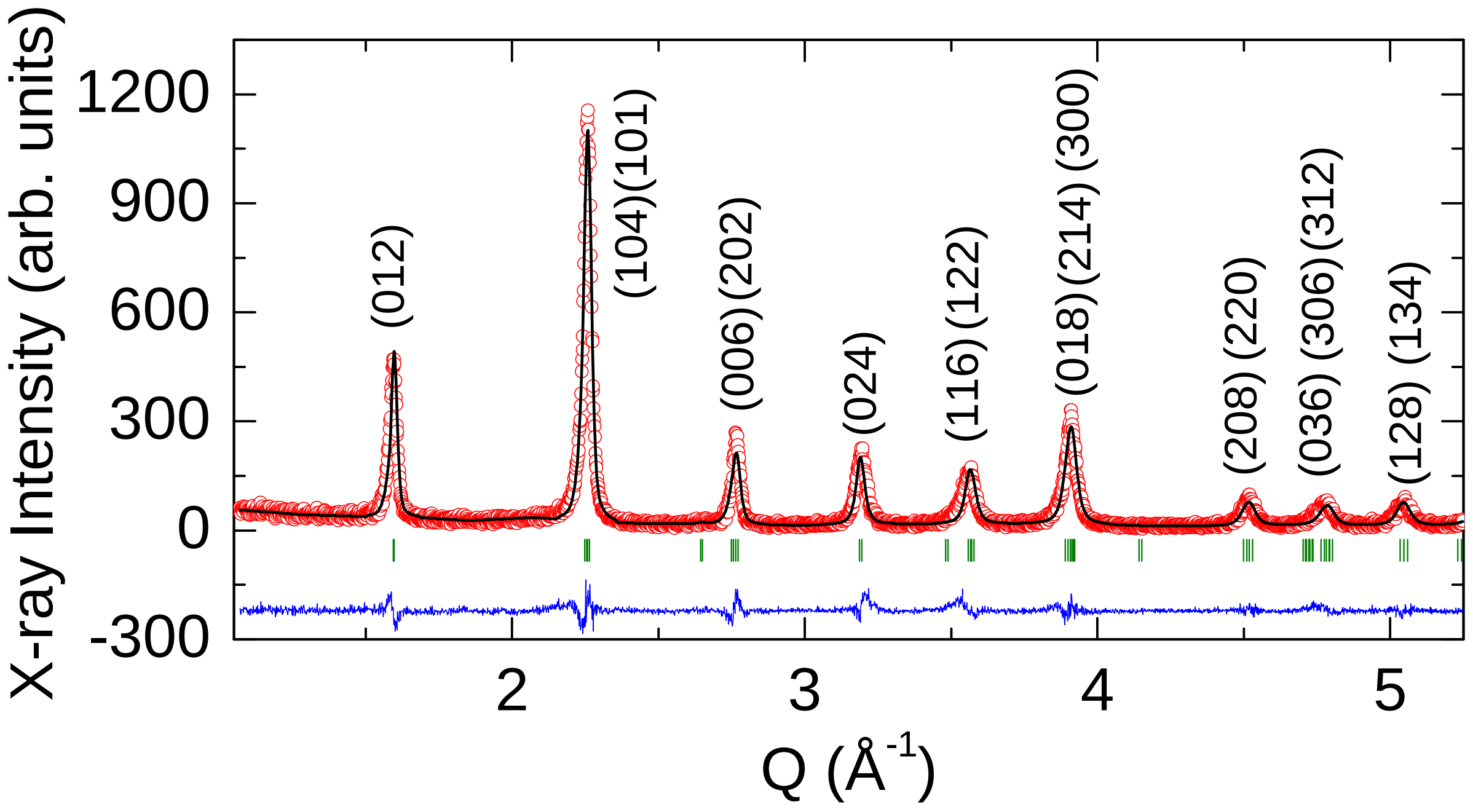

Fig. 1

Manna *et al.*

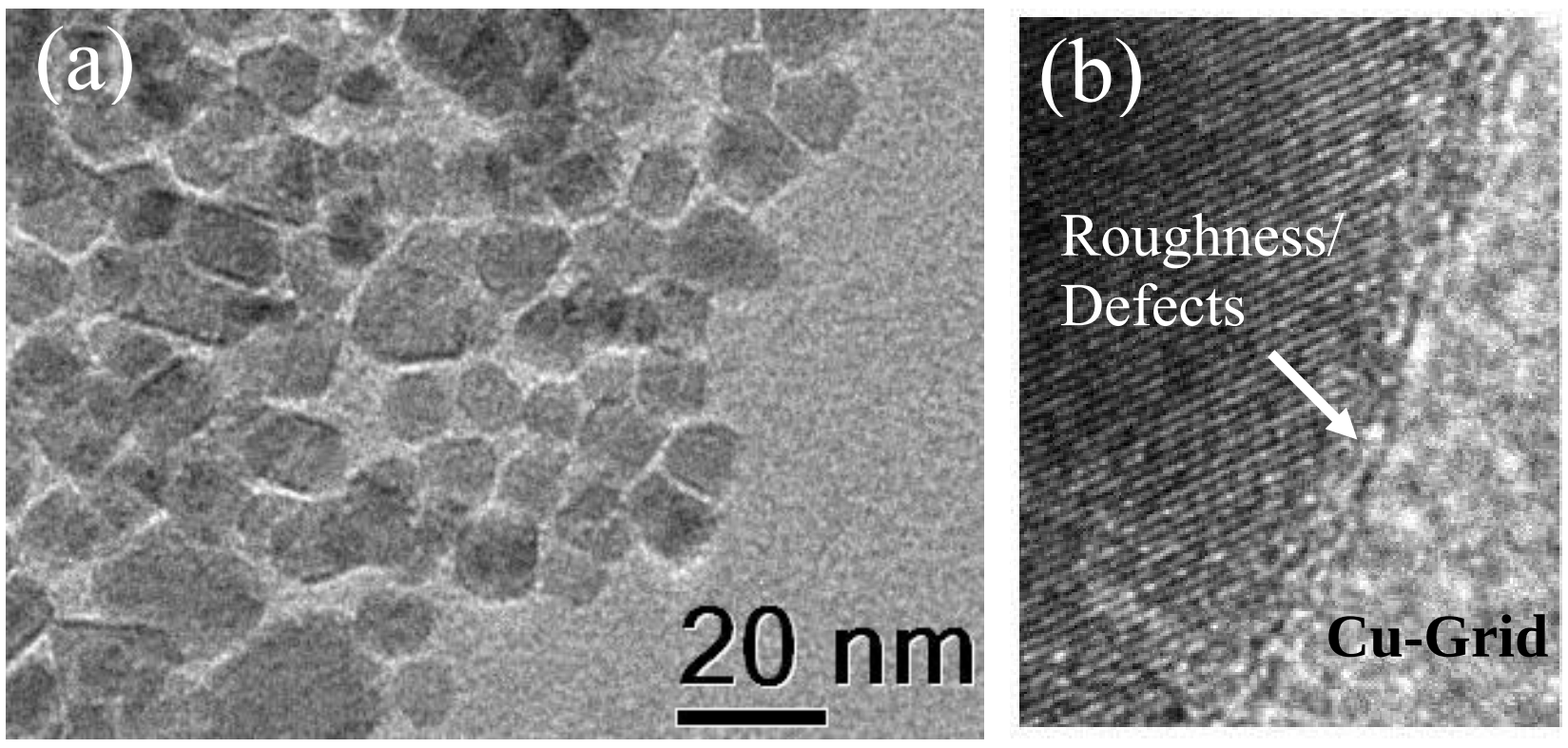

Fig. 2

Manna *et al.*

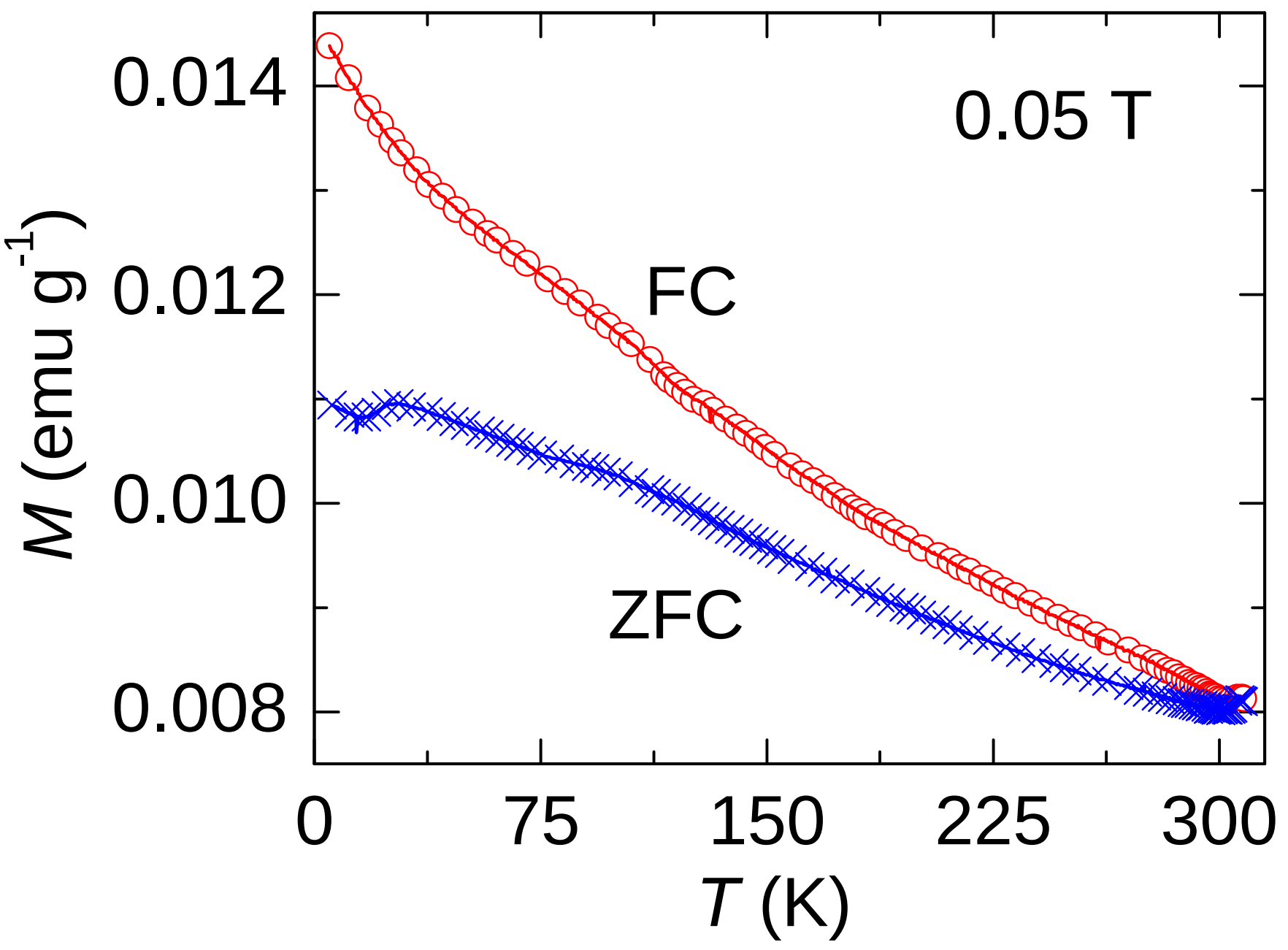

Fig. 3

Manna *et al.*

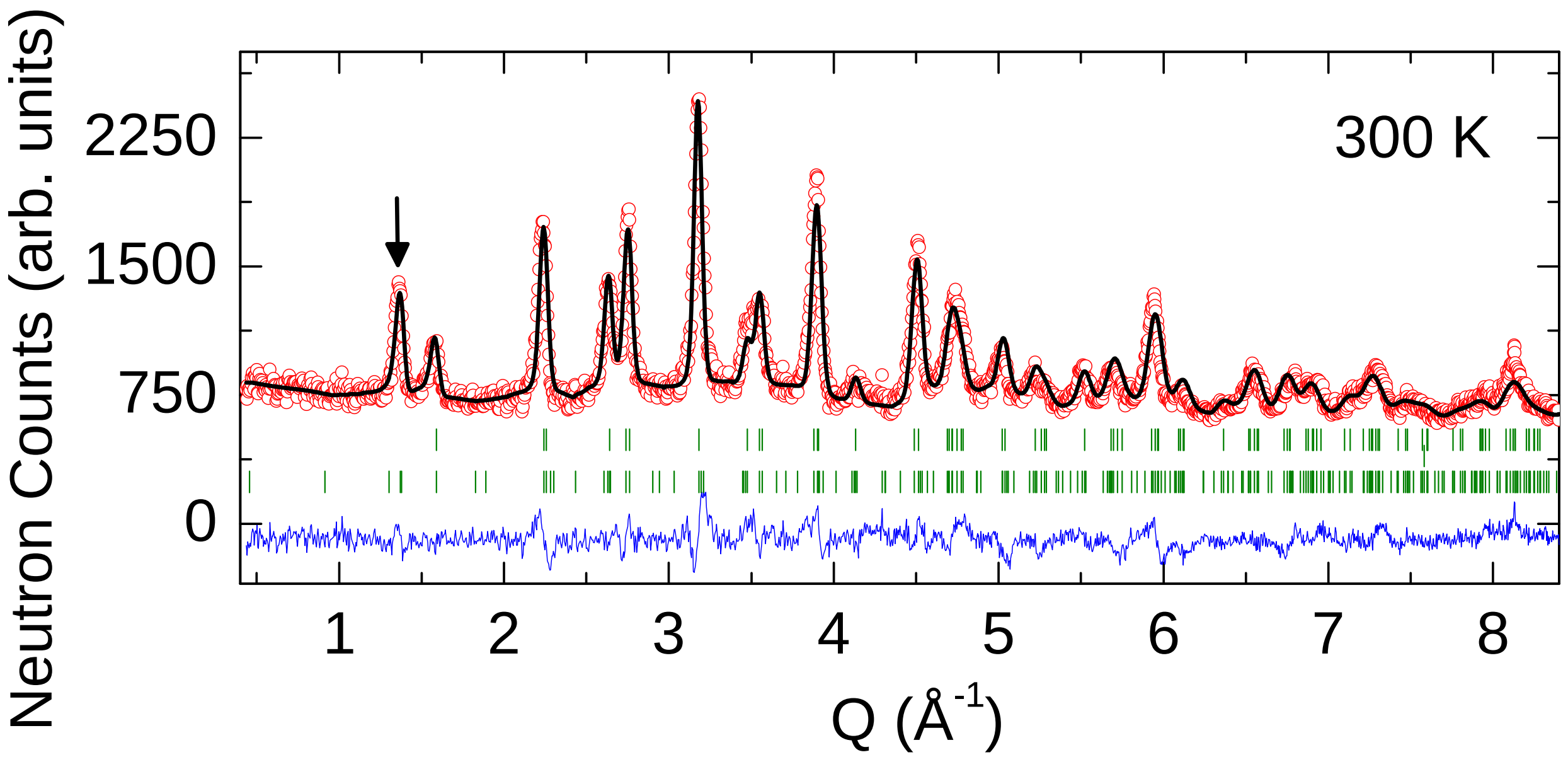

Fig. 4

Manna *et al.*

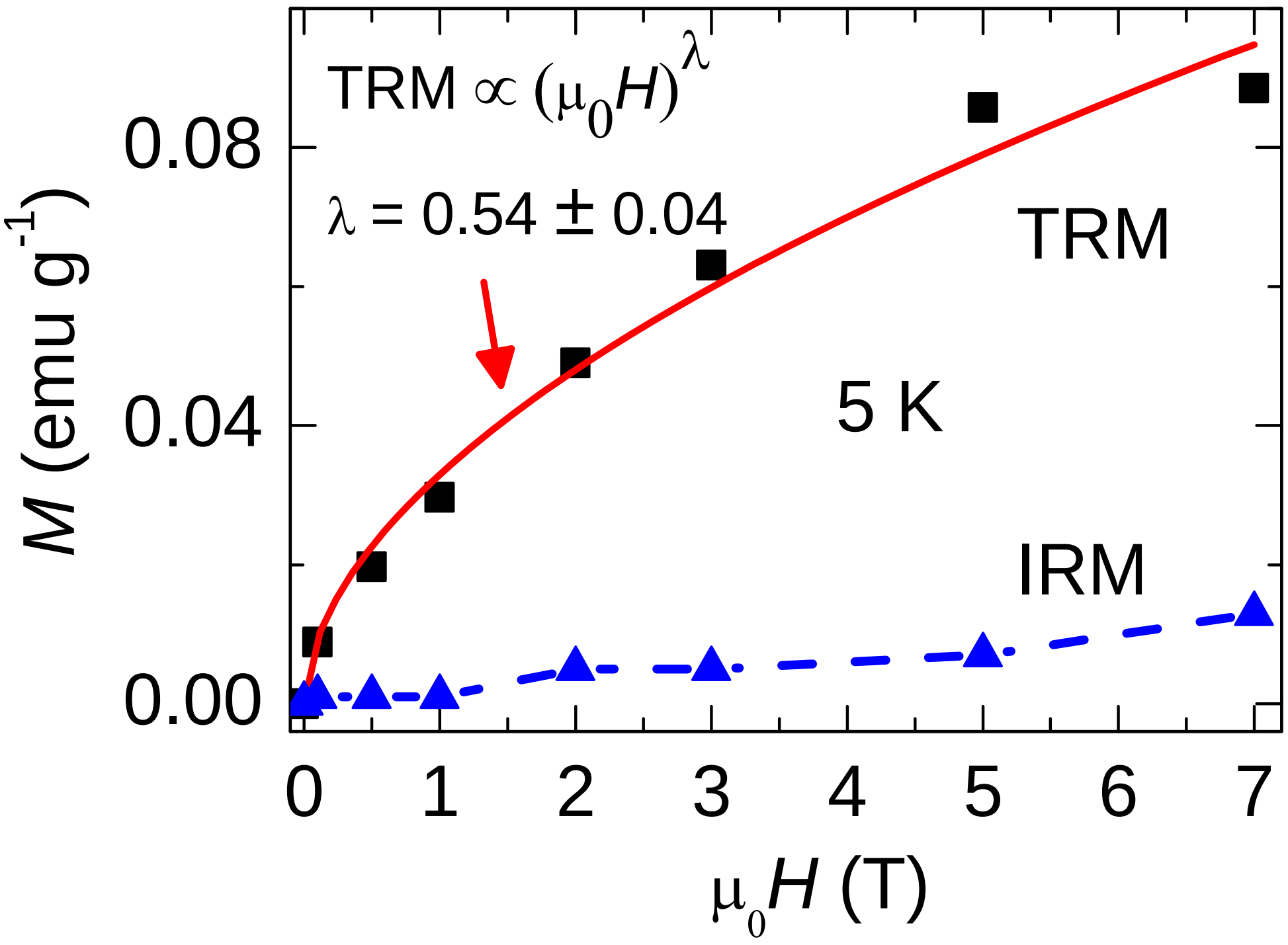

Fig. 5

Manna *et al.*

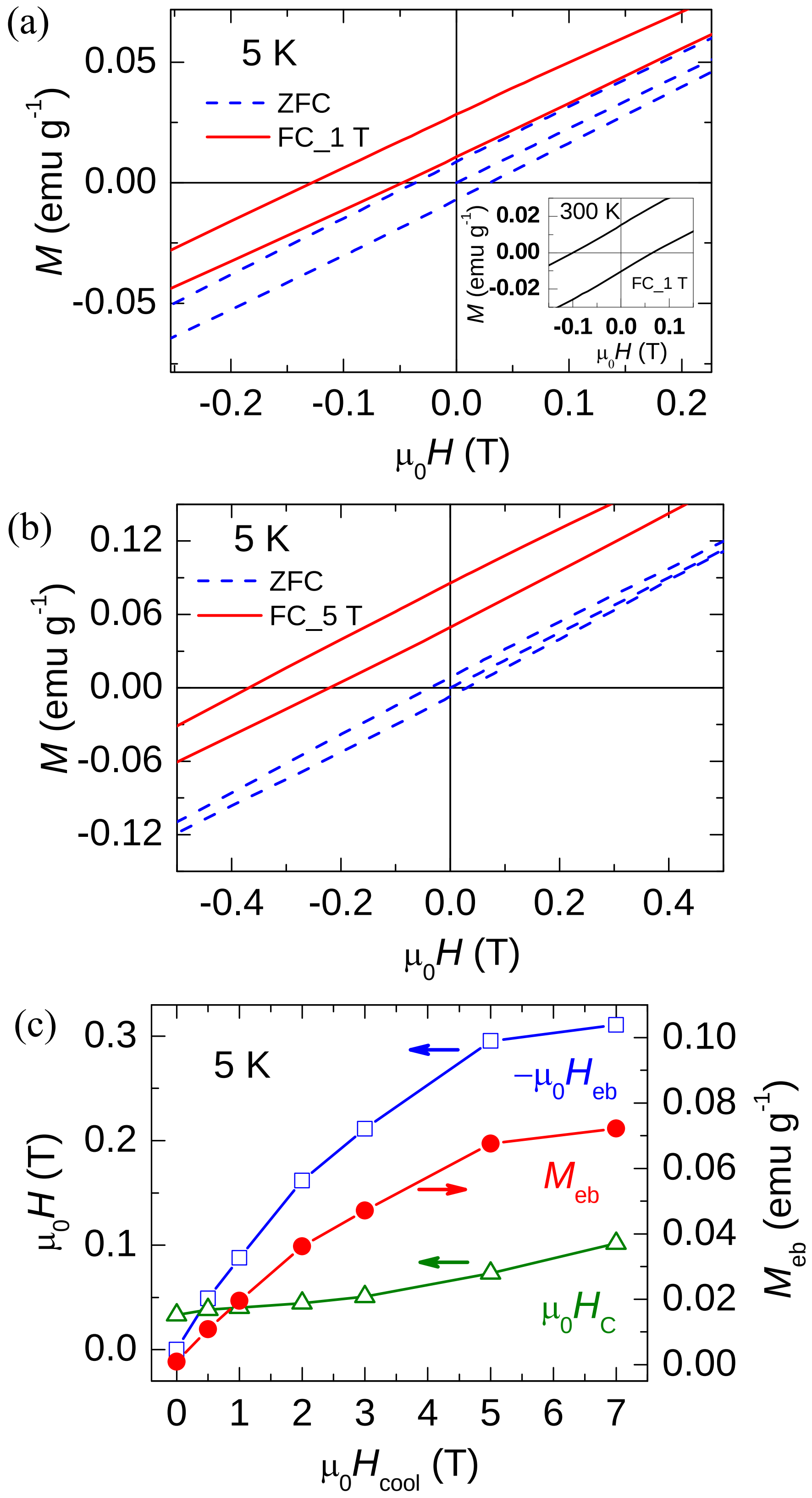

Fig. 6

Manna *et al.*

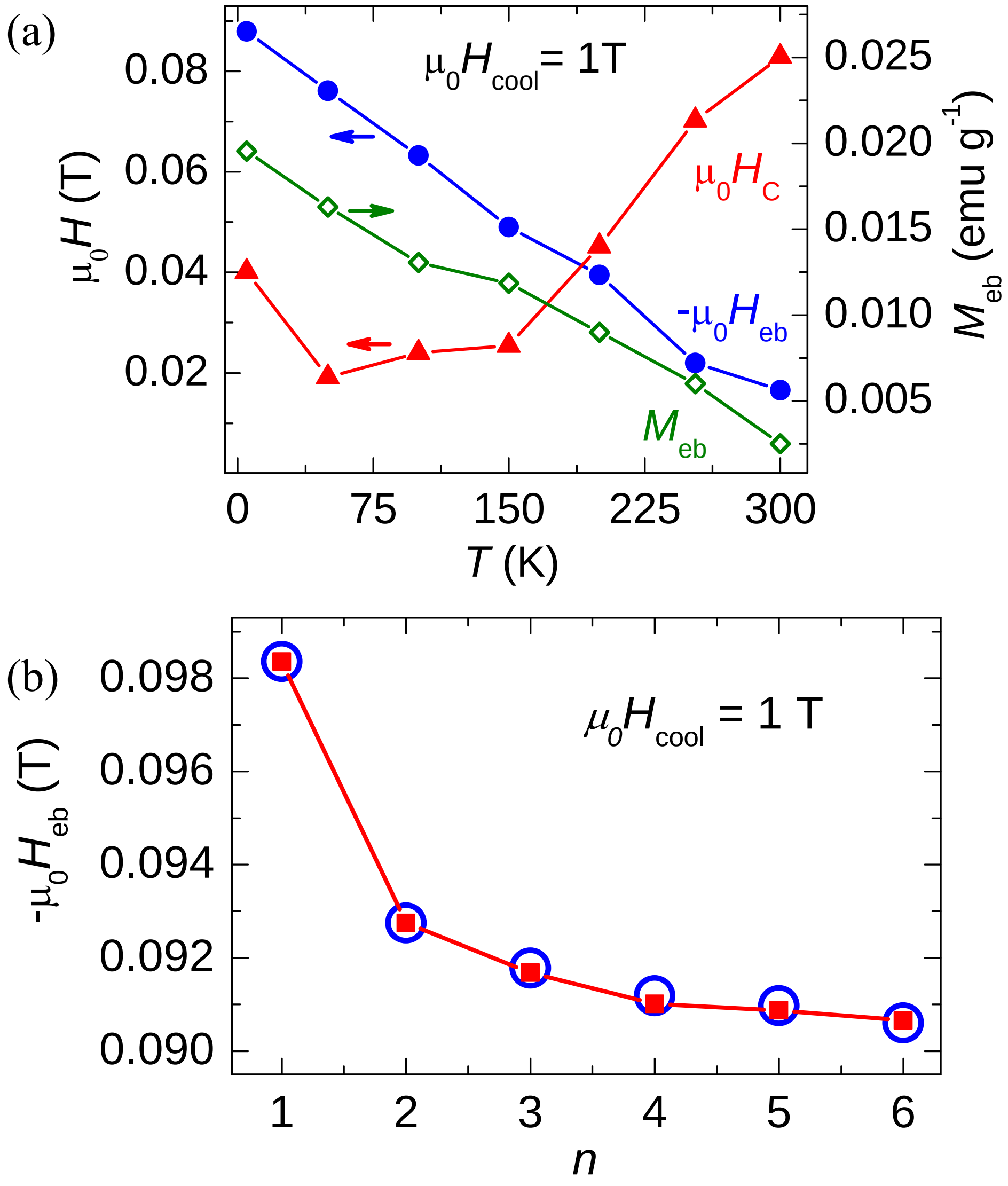

Fig. 7

Manna *et al.*